\documentclass{jfm}
\usepackage{amsmath,amssymb}
\usepackage{graphicx,rotating,booktabs}
\usepackage{natbib}

\usepackage{color}

\title[Transition to geostrophic convection: the role of the boundary conditions]{Transition to geostrophic convection: the role of the boundary conditions}
\author[R.P.J. Kunnen, R. Ostilla-M\'onico, E.P. van der Poel, R. Verzicco and D. Lohse]{Rudie P.J. Kunnen$^1$, Rodolfo Ostilla-M\'onico$^2$, Erwin P. van der Poel$^2$, Roberto Verzicco$^{3,2}$ and Detlef Lohse$^2$}

\affiliation{$^1$ Fluid Dynamics Laboratory, Department of Applied Physics and J.M. Burgers Centre for Fluid Dynamics, Eindhoven University of Technology, P.O. Box 513, 5600 MB Eindhoven, The Netherlands \\ $^2$ Physics of Fluids Group, Mesa+ Institute and J.M. Burgers Centre for Fluid Dynamics, University of Twente, P.O. Box 217, 7500 AE Enschede, The Netherlands \\  $^3$ Dipartimento di Ingegneria Industriale, University of Rome ``Tor Vergata", Via del Politecnico 1, Roma 00133, Italy}

\pubyear{}
\volume{}
\pagerange{}
\date{\today}

\begin{document}
\maketitle

\begin{abstract}
Rayleigh--B\'enard (RB) convection, the flow in a fluid layer heated from below and cooled from above, is used to analyze the transition to the geostrophic regime of thermal convection. In the geostrophic regime, which is of direct relevance to most geo- and astrophysical flows, the system is strongly rotated while maintaining a sufficiently large thermal driving to generate turbulence. We directly simulate the Navier--Stokes equations for two values of the thermal forcing, i.e. $Ra=10^{10}$ and $Ra=5\cdot10^{10}$,  a constant Prandtl number~$Pr=1$, and vary the Ekman number in the range $Ek=1.3\cdot10^{-7}$ to $Ek=2\cdot10^{-6}$ which satisfies both requirements of super-criticality and strong rotation. We focus on the differences between the application of no-slip vs. stress-free boundary conditions on the horizontal plates. The transition is found at roughly the same parameter values for both boundary conditions, i.e. at~$Ek\approx 9\times 10^{-7}$ for~$Ra=1\times 10^{10}$ and at~$Ek\approx 3\times 10^{-7}$ for~$Ra=5\times 10^{10}$. However, the transition is gradual and it does not exactly coincide in~$Ek$ for different flow indicators. In particular, we report the characteristics of the transitions in the heat transfer scaling laws, the boundary-layer thicknesses, the bulk/boundary-layer distribution of dissipations and the mean temperature gradient in the bulk. The flow phenomenology in the geostrophic regime evolves differently for no-slip and stress-free plates. For stress-free conditions the formation of a large-scale barotropic vortex with associated inverse energy cascade is apparent. For no-slip plates, a turbulent state without large-scale coherent structures is found; the absence of large-scale structure formation is reflected in the energy transfer in the sense that the inverse cascade, present for stress-free boundary conditions, vanishes.
\end{abstract}

\section{Introduction}
Natural convection is ubiquitous in Nature. It is found not only in the Earth's interior and oceans, but also in planetary atmospheres and also inside stars \citep{ms99,m00,rg00,haw05}. In all of those flows, the background rotation induces a Coriolis force, which significantly affects the system, changing not only the flow phenomenology but also the heat transport and the amount of mixing of different species. 

Rotating Rayleigh--B\'enard (RB) convection, the flow between two rotating parallel plates heated from below and cooled from above, is commonly used as a model for studying rotating thermal convection. Rotating a RB system induces many changes. In non-rotating RB, the flow consists mainly of plume-like structures. With increasing rotation, these plumes give way to columnar vortices \citep{r69,zes93,scl13}. For sufficiently large rotation, and also sufficiently low viscosity, the flow becomes quasi-two-dimensional (Q2D) owing to the Taylor--Proudman theorem \citep{g68}. This theorem implies that all slow, large-scale motions become two-dimensional (2D), i.e. independent of the direction parallel to the axis of rotation. However, fast, small-scale three-dimensional (3D) dynamics persists. The large-scale motions exhibit the so-called geostrophic balance between the pressure gradient and the Coriolis force \citep{g68}. Thus, this strongly-rotating regime of natural convection has been referred to as `geostrophic turbulence' \citep{sjkw06,jkrv12,jrgk12}. In particular, the name ``geostrophic turbulence'' has been introduced to refer to flow phenomenology as a state where vertical coherence of the flow has been lost \citep{jrgk12}. This is in contrast with the vertically aligned vortical plumes or the convective Taylor columns with strong vertical correlation that characterize the classical regime. The geostrophic regime, on which we focus here, has revealed distinctly different properties compared to the `classical' turbulent rotating convection with columnar vortices.  For a recent review on the classical regime, we refer the reader to \cite{scl13}. 


The difficulty of achieving the geostrophic regime is twofold.  Rotation not only changes the flow topology, but it also stabilizes the flow. \cite{c61} demonstrated using linear stability analysis that convective instability sets in at increasingly higher temperature differences when rotation is applied, or, in other words, the critical value of the Rayleigh number (the non-dimensionless temperature difference between the plates) rises as a function of the non-dimensional rotation rate, i.e. the Ekman number. Both a high rotation rate and a significant level of thermal driving, to remain turbulent even with respect to the increased critical Rayleigh number, are needed to achieve the geostrophic regime. Most earlier experimental and numerical studies of rotating RB \citep{r69, zes93, jlmw96, le97, ve02, kcg08epl, kcg08prl, le09, zscvla09, ksnha09, st09, st10, za10, wszcla10, scl10, kgc10jfm, ksoshc11, le11, wa11a, wa11b, scl12, kcc13, hs14} have not conclusively ventured deep into the geostrophic regime. The distinction between the geostrophic regime and the classical regime has been based on the results of numerical simulations \citep{sjkw06,jrgk12} which use a reduced set of the Navier--Stokes equations describing convection in the asymptotic limit of rapid rotation as a function of the forcing parameter~$\widetilde{Ra}=RaEk^{4/3}$, where~$Ra$ and~$Ek$ are the Rayleigh and Ekman numbers, respectively, to be defined later. Using the reduced equations, \citet{jkrv12} revealed the scaling laws relating heat transfer and $\widetilde{Ra}$ in the geostrophic regime, which were distinctly different from those previously reported for other flow regimes. Recent experiments by \cite{en14,csrgka15}, capable of simultaneously achieving very high~$Ra\gtrsim10^9$ and very low~$Ek\lesssim10^{-6}$ have also found the transitions in the scaling laws for the heat transfer reported previously.

Using the reduced equations, \citet{rjkw14} have shown an inverse energy cascade in the geostrophic regime. In an inverse cascade, unlike the regular cascade of 3D homogeneous isotropic turbulence, energy flows from the small length scales to larger ones. This leads to the formation of large-scale structures that typically become comparable in size to the domain in which they reside. Such self-organisation has also been reported in recent direct numerical simulations (DNSs) employing the full Navier--Stokes equations and stress-free boundaries \citep{fsp14,ghj14}. The reduced equations automatically imply taking stress-free boundaries which excludes the no-slip-style boundaries with their associated Ekman boundary layers \citep{g68}. \cite{sljvcrka14} made the step to full Navier--Stokes simulations in the geostrophic regime with no-slip boundaries. The initial findings include the absence of vortex condensation and a higher heat flux for no-slip plates than for stress-free at the same parameter values \citep{st10}. From a theoretical point of view, the geostrophic regime has also received some attention. \cite{e15} used weakly nonlinear theory to explain the scaling laws relating heat transfer and driving near the onset of convection.
 
However, a complete picture on what is happening during the transition to the geostrophic regime, and where it takes place in the parameter space is still missing. In this paper we present numerical simulations covering the transition to geostrophic convective turbulence using the full Navier--Stokes equations for a single Prandtl number. We analyse in detail the effects of the choice of boundary conditions, i.e. including or omitting the Ekman layers. In section~\ref{ch:num_setup} we describe the numerical method and give the parameter values for the runs. The results for the convective heat transfer are presented in section~\ref{ch:nusselt}. In section~\ref{ch:dissipation} we consider the effects of rotation on the boundary layer scales and on the volumetric distribution of kinetic-energy and thermal-variance dissipation, an approach which has allowed for the Grossmann--Lohse theory of heat transfer in non-rotating RB flow \citep{gl00,gl01,gl04,ste13}. The flow phenomenology and its relation with the spectral energy transfer is considered in section~\ref{ch:flow_phenom}. We conclude with an interpretation and discussion of these findings in section~\ref{ch:discussion}.

\section{Simulation details\label{ch:num_setup}}

We have conducted a set of direct numerical simulations (DNS) of 3D rotating RB in a horizontally periodic Cartesian computational box. By using a second-order energy-conserving, finite-difference code with fractional time-stepping \citep{vo96}, we march in time the Navier--Stokes equations plus an advection-diffusion equation for temperature, with the usual Boussinesq approximations~\citep{c61}:

\begin{equation}
\frac{\partial\boldsymbol{u}}{\partial t} + (\boldsymbol{u \cdot\nabla}) \boldsymbol{u} + \frac{1}{Ro} \boldsymbol{e}_z \boldsymbol{\times u}= -\boldsymbol{\nabla} p + \sqrt{\frac{Pr}{Ra}} \nabla^2\boldsymbol{u} + \theta \boldsymbol{e}_z\, ,
\label{eq:ns}
\end{equation}

\begin{equation}
\frac{\partial\theta}{\partial t} + (\boldsymbol{u \cdot\nabla}) \theta = \frac{1}{\sqrt{RaPr}} \nabla^2\theta\, ,
\label{eq:heat}
\end{equation}

\noindent with the incompressibility constraint 

\begin{equation}
\boldsymbol{\nabla \cdot u} = 0\, ,
\label{eq:incomp}
\end{equation}

\noindent where~$\boldsymbol{u}$ is the velocity vector, $t$ is time, $\boldsymbol{e}_z$ is the unit vector in the vertical direction, $Ra$ is the Rayleigh number, i.e. the non-dimensional temperature difference, defined as  $Ra = g\beta\Delta L^3 / (\nu\kappa)$  with $L$ the height of the system, $\beta$ the thermal expansion coefficient of the fluid, $g$ the gravitational acceleration, $\Delta$ the temperature difference between the bottom and top plates, and $\nu$ and $\kappa$ the kinematic viscosity and thermal diffusivity of the fluid, respectively. $Ro$ is the Rossby number, i.e. the inverse rotation rate, defined as $Ro = \sqrt{\beta g \Delta/L}/(2\Omega)$, where $\Omega$ is the angular rotation rate, $Pr$ is the Prandtl number of the fluid, $Pr = \nu/\kappa$ and $\theta$ the non-dimensional temperature. The equations (\ref{eq:ns})--(\ref{eq:incomp}) are non-dimensionalized by using $L$, $\Delta$, and the so-called free-fall velocity scale $\sqrt{\beta g\Delta L}$. Note that centrifugal buoyancy is neglected here; this means that we are implicitly making the customary assumption that the Froude number~$Fr=\Omega^2 R/g \ll 1$ \citep{scl13}, where~$R$ is the horizontal distance to the rotation axis, i.e. the radius of the cylinder in most experiments and simulations. We also define the Nusselt number, i.e. the non-dimensional heat transfer as $Nu=(\langle u_z \theta\rangle_{A,t}-\kappa\partial{\langle\theta\rangle_{A,t}})/(\kappa\Delta L^{-1})$, with $\langle ... \rangle_{A,t}$ representing the averaging operator in time and also spatially over a horizontal plane.

The explored parameter values are given in table~\ref{ta:numsims}. We vary the rotation rate~$Ro$ at a constant thermal driving~$Ra$ for two values of $Ra$, while fixing the Prandtl number to~$Pr=1$. The Ekman number, defined as~$Ek=\nu/(2\Omega L^2)=Ro\sqrt{Pr/Ra}$, is small enough to enter into the geostrophic regime, as can be seen from the table. For completeness, we also define the Taylor number $Ta=Ek^{-2}$. The aspect-ratio, $\Gamma = D/L$, where $D$ is the simulation box periodicity length in the horizontal directions, is set to ten times the most unstable wavelength for convective instability~$L_c$, i.e. $\Gamma=10L_c$. $L_c$ scales asymptotically as~$L_c=4.82 Ek^{1/3}$ with minor corrections at finite~$Ek$ \citep{c61,nb65}. Here, we just take $\Gamma=48.2 Ek^{1/3}$. The boundary conditions for temperature are fixed as~$\theta=1$ at the bottom plate and~$\theta=0$ at the top plate. For velocity we employ both no-slip (NS) boundary conditions, i.e. $\boldsymbol{u}=\boldsymbol{0}$ at both plates, and stress-free (SF) boundary conditions, i.e. $\partial_z u_x=\partial_z u_y=0$ and~$u_z=0$ at the plates.

\begin{table}
\centering
\begin{tabular}{cccccccc}
$Ra$              & $Ek$                 & $Ro$    & $\widetilde{Ra}$ & $\Gamma$ & $N_x\times N_y\times N_z$  & $Nu_\mathrm{SF}$ & $Nu_\mathrm{NS}$ \\
 & & & & & & & \\
$1\times 10^{10}$ & $4.00\times 10^{-7}$ & $0.040$ & $29.5$           & $0.36$   & $384\times 384\times 768$  & $8.82$           & $21.0$ \\
$1\times 10^{10}$ & $4.00\times 10^{-7}$ & $0.040$ & $29.5$           & $0.71$   & $768\times 768\times 768$  & $9.13$           & $21.0$ \\
$1\times 10^{10}$ & $6.00\times 10^{-7}$ & $0.060$ & $50.6$           & $0.41$   & $384\times 384\times 768$  & $20.7$           & $31.4$ \\
$1\times 10^{10}$ & $9.00\times 10^{-7}$ & $0.090$ & $86.9$           & $0.46$   & $384\times 384\times 768$  & $46.2$           & $50.2$ \\
$1\times 10^{10}$ & $1.20\times 10^{-6}$ & $0.12$  & $127.5$          & $0.51$   & $384\times 384\times 768$  & $68.5$           & $65.2$ \\
$1\times 10^{10}$ & $1.50\times 10^{-6}$ & $0.15$  & $171.7$          & $0.55$   & $384\times 384\times 768$  & $91.0$           & $76.0$ \\
$1\times 10^{10}$ & $2.00\times 10^{-6}$ & $0.20$  & $252.0$          & $0.61$   & $512\times 512\times 768$  & $113.7$          & $83.5$ \\
 & & & & & & & \\
$5\times 10^{10}$ & $1.34\times 10^{-7}$ & $0.030$ & $34.3$           & $0.25$   & $512\times 512\times 1024$ & $9.20$           & $21.1$ \\
$5\times 10^{10}$ & $1.79\times 10^{-7}$ & $0.040$ & $50.4$           & $0.27$   & $512\times 512\times 1024$ & $18.2$           & $30.8$ \\
$5\times 10^{10}$ & $2.95\times 10^{-7}$ & $0.066$ & $98.3$           & $0.32$   & $512\times 512\times 1024$ & $52.9$           & $61.5$ \\
$5\times 10^{10}$ & $4.02\times 10^{-7}$ & $0.090$ & $148.6$         & $0.36$   & $512\times 512\times 1024$ & $95.0$           & $88.3$ \\
$5\times 10^{10}$ & $4.92\times 10^{-7}$ & $0.11$  & $194.2$          & $0.38$   & $512\times 512\times 1024$ & $117.0$          & $103.5$ \\
$5\times 10^{10}$ & $6.71\times 10^{-7}$ & $0.15$  & $293.6$          & $0.42$   & $512\times 512\times 1024$ & $159.6$          & $119.5$ \\
\end{tabular}
\caption{\label{ta:numsims}Parameter values for the computations. For all runs~$Pr=1$. Each parameter set has been run with both NS and SF boundary conditions. Included are: Rayleigh number~$Ra$, Ekman number~$Ek$, Rossby number~$Ro$, $\widetilde{Ra}=RaEk^{4/3}$ \citep{sjkw06}, domain aspect ratio~$\Gamma$ and number of gridpoints~$N_x\times N_y\times N_z$ in the periodic directions and the vertical direction, respectively. We also list the resulting Nusselt numbers~$Nu_\mathrm{SF}$ and~$Nu_\mathrm{NS}$.}
\end{table}

We tested the dependence on~$\Gamma$ of the simulations by running two cases at~$Ra=1\times 10^{10}$ and~$Ro=0.04$ with a twice larger~$\Gamma$ (thus increasing the computational box size by a factor four, and hence the computational load by at least that amount). The Nusselt number is the same for no-slip plates but shows some difference in the stress-free case. For no-slip plates we do not expect a strong dependence on~$\Gamma$ as long as it is large enough. However, for stress-free plates differences may occur which are related to the flow structure. This will be addressed in more detail in Section~\ref{ch:flow_phenom}. 

\begin{figure}
\centerline{\includegraphics[width=0.49\textwidth,trim={0 0 3cm 0cm},clip]{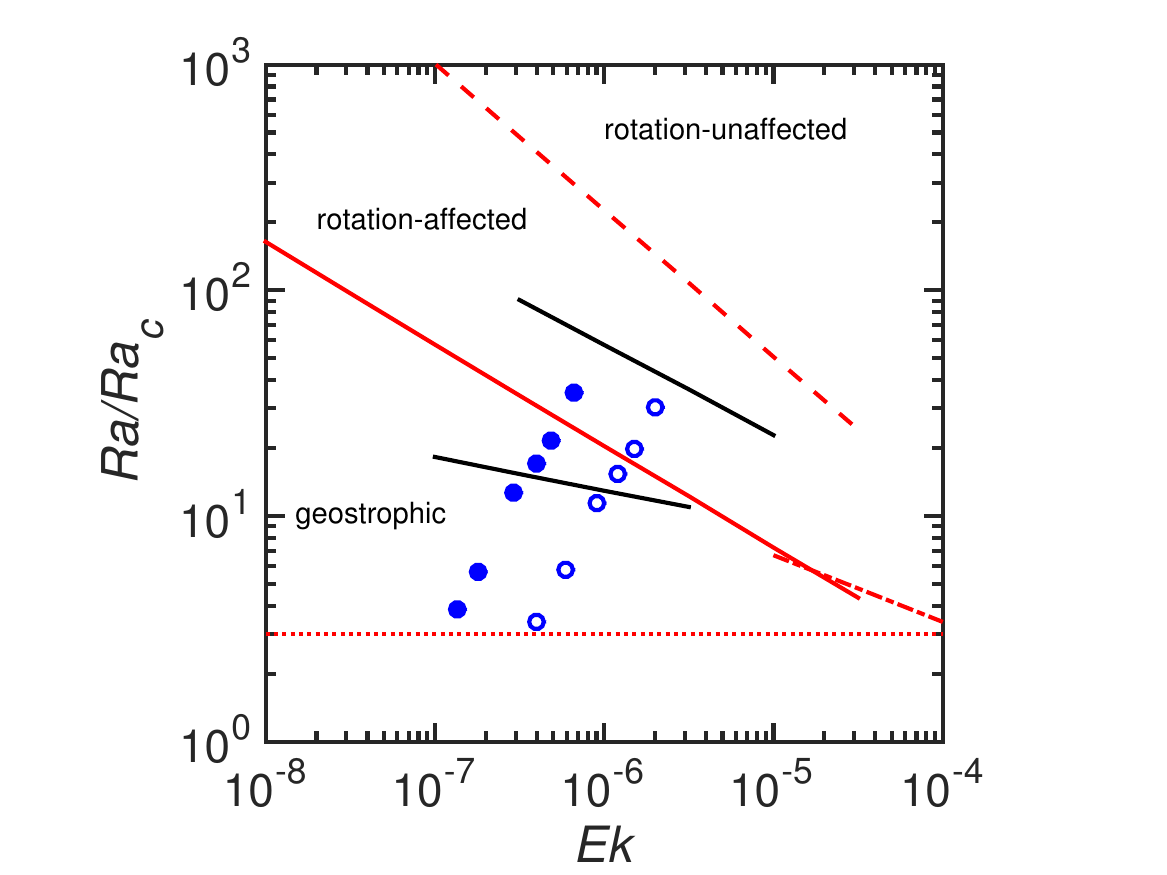}
\includegraphics[width=0.49\textwidth,trim={0 0 3cm 0cm},clip]{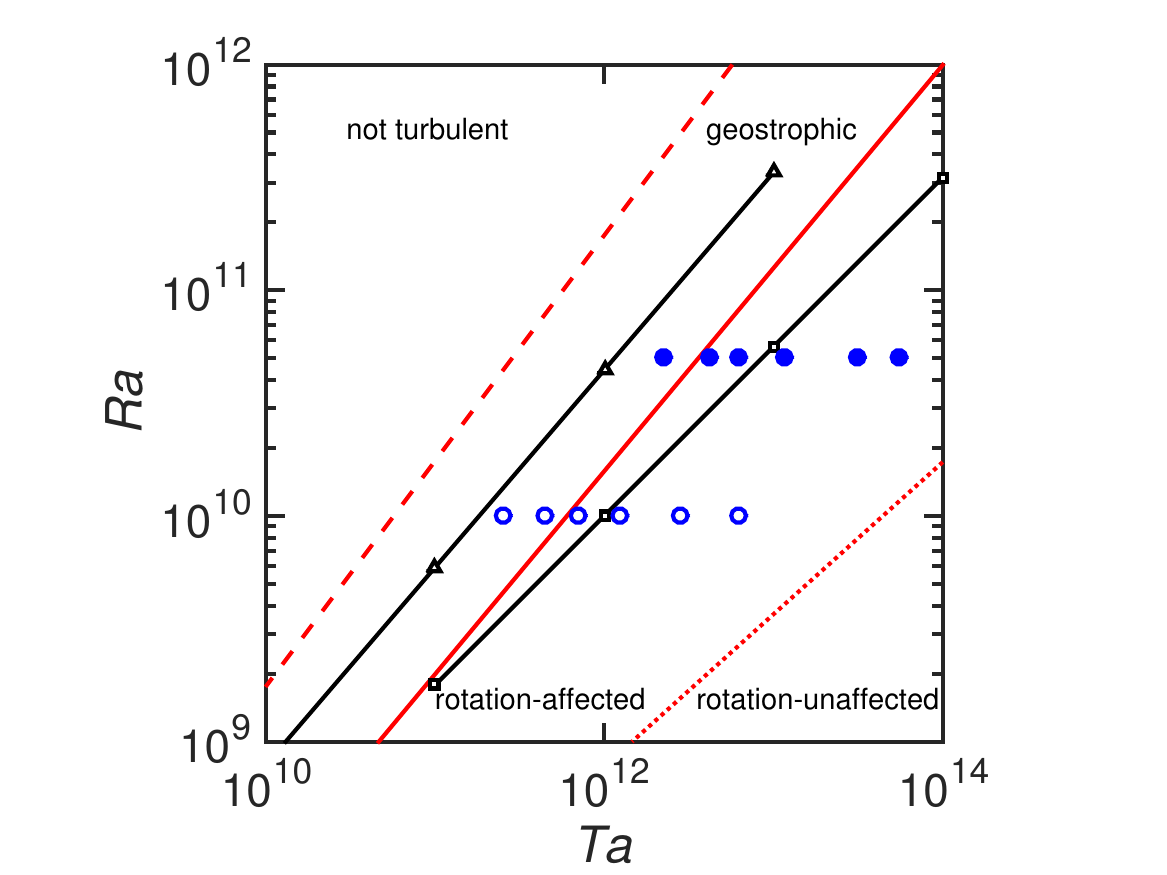}}
\caption{\label{fi:phase_diag} Phase diagram of rotating convection in the ($Ra/Ra_c$,$Ek$) parameter space, as suggested by \citet{en14}, and in the ($Ta$,$Ra$) parameter space. The blue circles indicate the NS/SF simulation pairs of this work, these are either open for $Ra=1\times 10^{10}$ or filled for $Ra=5\times 10^{10}$. Three different regimes of convection can be discerned: non-rotating convection, rotation-affected convection (classical rotating convection) and geostrophic convection. The lines display various relations suggested in the literature that bound the regimes. \citet{en14} suggested $Ra/Ra_c=3$ as a lower bound of the geostrophic regime (dotted red line), $Ra=0.25Ek^{-1.8}$ for the transition between the geostrophic and rotation-affected regimes, and $Ro=0.35$ for the transition to the non-rotating regime. The dash-dotted red line is a transition valid for higher $Pr\approx 6$ \citep{en14}. Two alternative predictions for the transition to the geostrophic regime are also displayed: $Ra=1.4Ek^{-7/4}$ \citep[][black solid line with triangles]{ksnha09} and $Ra\approx 10Ek^{-3/2}$ \citep[][black solid line with squares]{ksa12}.}
\end{figure}

To indicate how the current simulations fit in with the previous work on this topic, we display our current parameter values in the ($Ek$, $Ra/Ra_c$) phase diagram of figure~\ref{fi:phase_diag}, where $Ra_c$ is the critical Rayleigh number for the onset of the convective instability with rotation, i.e. $Ra_c=8.7Ek^{-4/3}$ \citep{c61}, and in the ($Ta$,$Ra$) phase diagram. The left panel of this figure is based on figure 4 of \citet{en14}, and shows how the parameter values of the simulations in this manuscript are positioned relative to some of the proposed bounds on the geostrophic regime. Typically, a lower bound for the geostrophic regime is chosen such that the resulting flow is supercritical enough for a turbulent flow to develop. We follow \citet{en14} by choosing $Ra/Ra_c=3$ (dotted red line in figure~\ref{fi:phase_diag}). \cite{en14} discern two additional transitions based on their heat transfer measurements in cryogenic helium with $Pr=0.7$. When reducing the Rayleigh number at constant~$Ek$, the first transition seen is when rotation starts to reduce the heat transfer. This is well-described by~$Ro=0.35$ (dashed red line in figure~\ref{fi:phase_diag}). When $Ra$ is reduced even further, a transition to a steeper scaling law relating $Nu$ and $Ra$ was found. \citet{en14} interpreted this as the transition to the geostrophic regime. This transition was best described by the relation $Ra=0.25 Ek^{-1.8}$ (solid red line in figure~\ref{fi:phase_diag}) according to their data. For completeness, two other suggested relations for the transfer to the geostrophic regime have also been included in figure~\ref{fi:phase_diag}: $Ra=1.4Ek^{-7/4}$ as suggested by \citet{ksnha09} (black solid line with triangles) and $Ra\approx 10Ek^{-3/2}$ from \citet{ksa12} (black solid line with squares). From the diagram, we expect our simulations to show a transition from the rotation-affected to the geostrophic regime for all criteria, except the one suggested by \citet{ksnha09}.

\section{Heat transfer\label{ch:nusselt}}

In this section, we investigate the convective heat transfer through the fluid layer as a function of the applied control parameters. In the geostrophic regime, no consensus has been reached on the heat-transfer dependence~$Nu(Ra,Pr,Ek)$, in particular because it is challenging to achieve the extreme parameter values for~$Ra$ and~$Ek$ in both experiments and simulations. We summarize the results from the literature reported earlier in table~\ref{ta:nu_scaling}, and indicate the method (experimental, numerical or from theory) and range of parameters considered. It must be emphasized that most of these works are outside of the geostrophic regime; the exceptions are the theories by \citet{ksa12} and \citet{jkrv12}, as well as the numerical simulations by the latter authors, which consider the asymptotically reduced equations for rapid rotation.

\begin{sidewaystable}
\vspace{15cm}
\centering
\begin{tabular}{ccccc}
Authors         & Method       & Scaling                   & Parameter range                                                & Remark \\
 & & & & \\
\citet{ksnha09} & exp. \& num. & $Nu\sim Ra^{6/5}Ek^{8/5}$ & $Ra < 6\times 10^9$; $1\times 10^{-6}\le Ek$; $1\le Pr\le 100$ & NS \\
 & & & & \\
\citet{st09}    & num.         & $Nu\sim Ra^{5/4}Ek^{3/2}$ & $Ra \le 2\times 10^8$; $Ek\ge 1\times 10^{-5}$ for~$Pr=0.7$    & SF \\
 & &                                                       & $Ra \le 1\times 10^8$; $Ek\ge 1\times 10^{-5}$ for~$Pr=7$      & SF\\
 & & & & \\
\citet{st10}    & num.         & $Nu\sim Ra^{5/4}Ek^{3/2}$ & $Ra \le 1\times 10^7$; $Ek\ge 1\times 10^{-4}$; $Pr=7$         & NS \\
 & & & & \\
\citet{ksa12}   & theor.       & $Nu\sim Ra^3Ek^4$         & $Ra\lesssim Ek^{-3/2}$                                         & \\
 & & & & \\
\citet{jkrv12}  & theor.       & $Nu-1\sim Ra^{3/2}Ek^2$   &                                                                & \\
                & num.         &                           & $0.3\le Pr\le 1$ for~$RaEk^{4/3}\gtrsim 80$                    & asymptotic eqs. \\
\end{tabular}
\caption{\label{ta:nu_scaling}Proposed and observed scaling laws for the heat transfer in rotation-dominated convection found in the literature, expressed as~$Nu(Ra,Ek)$. The Prandtl number dependence has been omitted as it is out of the scope for the present paper. We also mention the range of parameters for which the scalings are intended or observed, as well as the method by which the scalings have been obtained (`exp.' for experiments, `num.' for numerical simulations and `theor.' for scaling arguments). Finally, the boundary conditions for the numerical studies are mentioned. These are the topic of the current work. The simulations by \citet{jkrv12} are based on the asymptotic equations and employ what amounts to stress-free boundary conditions. Note that \citet{st09,st10} have defined the Ekman number without the factor~$2$, hence the different parameter values in this table compared to their papers.}
\end{sidewaystable}

The two recent experimental investigations which enter into the geostrophic regime show quite different results. \citet{en14} achieved~$4\times 10^9<Ra<4\times 10^{11}$ and~$2\times 10^{-7}<Ek<3\times 10^{-5}$ at~$Pr=0.7$ used cryogenic helium gas as the working fluid. Their data could be described as~$Nu\sim (Ra/Ra_c)^\gamma$, with~$\gamma\approx 1$ using direct measurement or~$1.2<\gamma <1.6$ after rescaling of the original data following existing theoretical arguments. The scaling ranges were not extensive enough to decisively discern between these scalings. On the other hand, \citet{csrgka15} employed water as the working fluid. In their tall, slender cell they could achieve~$1\times 10^{10}<Ra<1\times 10^{13}$ and~$2\times 10^{-8}< Ek<2\times 10^{-6}$ for~$3.5<Pr<6.5$. They also reported scaling as~$Nu\sim (Ra/Ra_c)^\gamma$, with a monotonically increasing~$\gamma$ from~$1.8$ at~$Ek=10^{-3}$ to~$3.6$ at~$Ek=10^{-7}$. However, at the highest rotation rates (lowest~$Ek$) there may be a significant effect of centrifugal buoyancy as at the sidewall the centrifugal acceleration can have a magnitude of up to~$40\%$ of gravity.

\begin{figure}
\centerline{\includegraphics[scale=0.5]{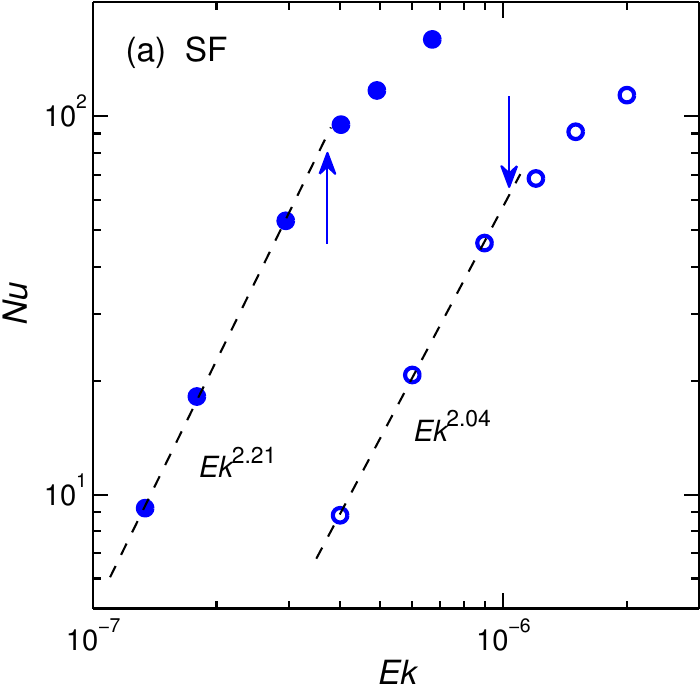}\ \ \includegraphics[scale=0.5]{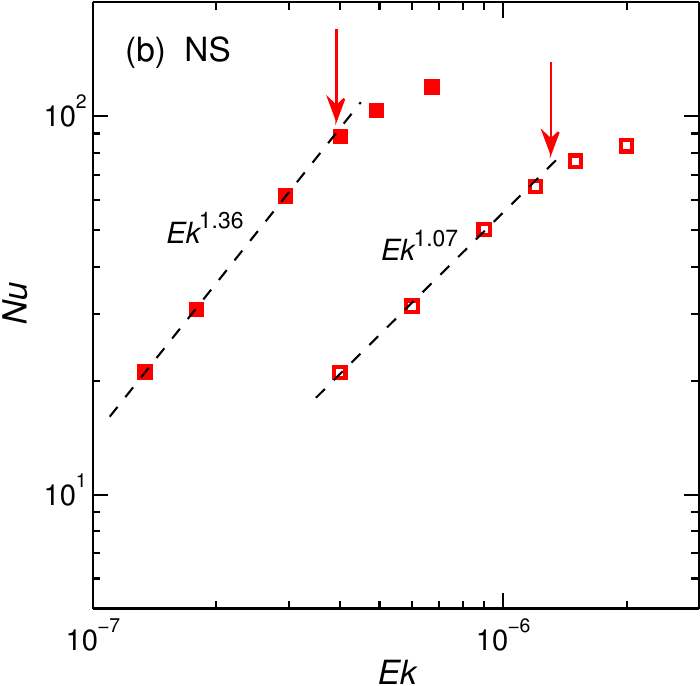}}
\caption{\label{fi:nusselt}Heat transfer (Nusselt number~$Nu$) as a function of the Ekman number~$Ek$. Open symbols are for~$Ra=1\times 10^{10}$; filled symbols for~$Ra=5\times 10^{10}$. The dashed lines depict fitted power-law slopes. The arrows indicate the transition points as we inferred it from these graphs. (a) SF plates. (b) NS plates.}
\end{figure}

Figure~\ref{fi:nusselt} shows $Nu$ as a function of $Ek$ obtained from the present simulations. By simple observation, it is clear that the boundary conditions (NS or SF) play a decisive role, even in the slope of the graph, i.e. the exponent $\alpha$ of the local scaling law $Nu\sim Ek^\alpha$. Both NS and SF boundary conditions display a transition in the scaling law, (indicated with arrows in the graph) at~$Ek\approx 9\times 10^{-7}$ for~$Ra=1\times 10^{10}$ and at~$Ek\approx 3\times 10^{-7}$ for~$Ra=5\times 10^{10}$, as evidenced by the slope change. This transition is generally considered the boundary between rotation-affected and rotation-dominated convection~\citep{en14}. At Ekman numbers below the transition we observe distinctly different scalings with~$Ek$. The SF exponents ($\alpha=2.04$ for~$Ra=1\times 10^{10}$ and~$\alpha=2.21$ for~$Ra=5\times 10^{10}$) match fairly well (especially for the lower $Ra$ case) with the theoretically predicted exponent~$\alpha=2$ of~\citet{jkrv12}, found to be valid for simulations of the reduced equations, with boundary conditions that can be described as stress-free. 

However, the NS runs reveal effective exponents~$\alpha=1.07$ for~$Ra=1\times 10^{10}$ and~$\alpha=1.36$ for~$Ra=5\times 10^{10}$, considerably lower than the SF runs and prounouncedly lower than the correlations inferred by~\citet{ksa12} which predict $\alpha=4$ for NS plates. \citet{st09,st10} have reported simulations with both NS and SF boundaries; they reported good agreement with the exponent~$1.5$, in fair agreement with the current NS runs (at least at~$Ra=5\times 10^{10}$) but somewhat low for SF. The experiments by \citet{en14} have provided approximate scaling exponents between~$1.6$ and~$2.1$, depending on the exact plotting convention to attain data collapse.

It is worth noting that there is quite a difference in the applied~$Ra$ between the various works. Typically, values of~$Ra$ up to~$5\times 10^9$ are applied in experiments and simulations~\citep{ksnha09,st09,st10,ksa12,ksb13}, while only recently higher values have been attained (c.f. simulations by \citet{sljvcrka14}, experiments by \citet{en14,csrgka15}, as well as the current simulations). It is plausible that what we observe is a new scaling regime opening up at such high~$Ra$, strongly affected by rapid rotation (very low~$Ek\sim O(10^{-7})$) but still vigorously turbulent (highly supercritical, i.e.~$Ra/Ra_c\gg 1$), i.e. the geostrophic regime. 

\begin{figure}
\centerline{\includegraphics[width=0.49\textwidth,trim={0 0 0 0}, clip]{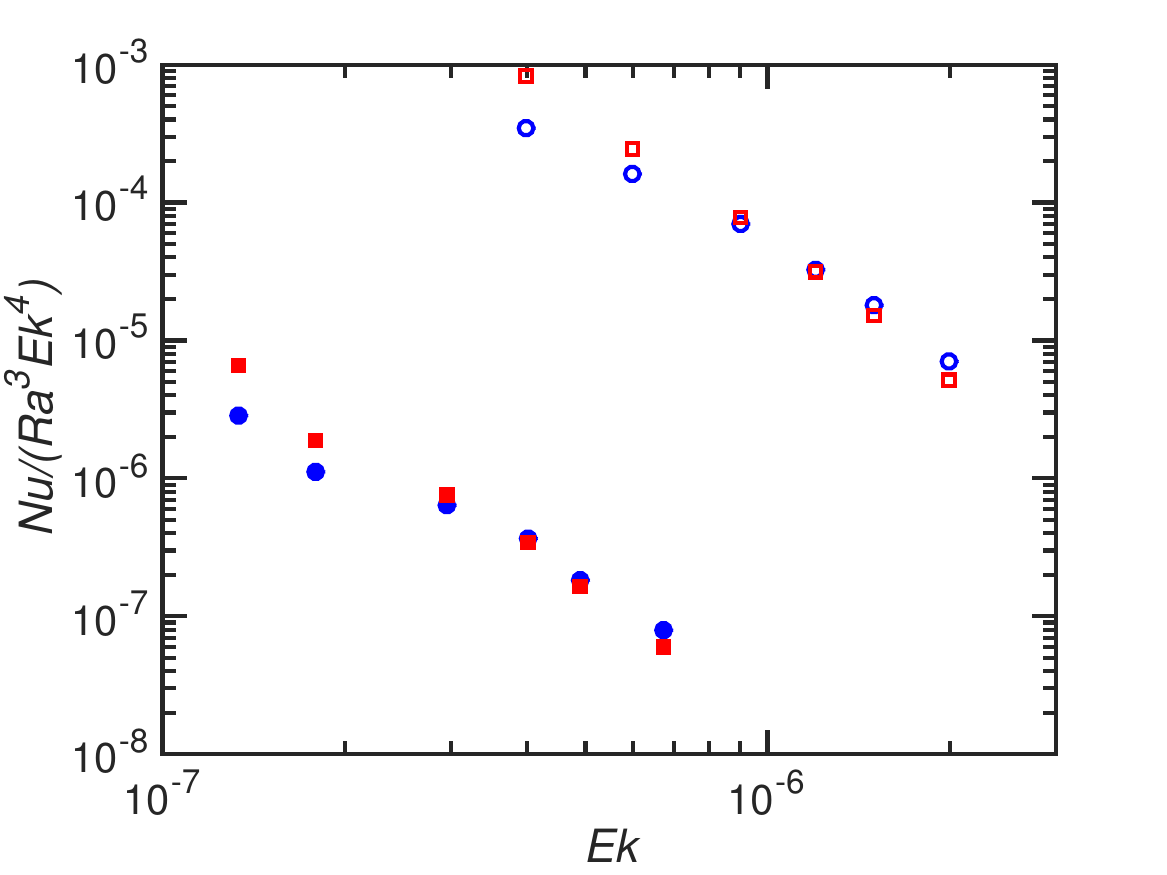}
\includegraphics[width=0.49\textwidth,,trim={0 0 0 0}, clip]{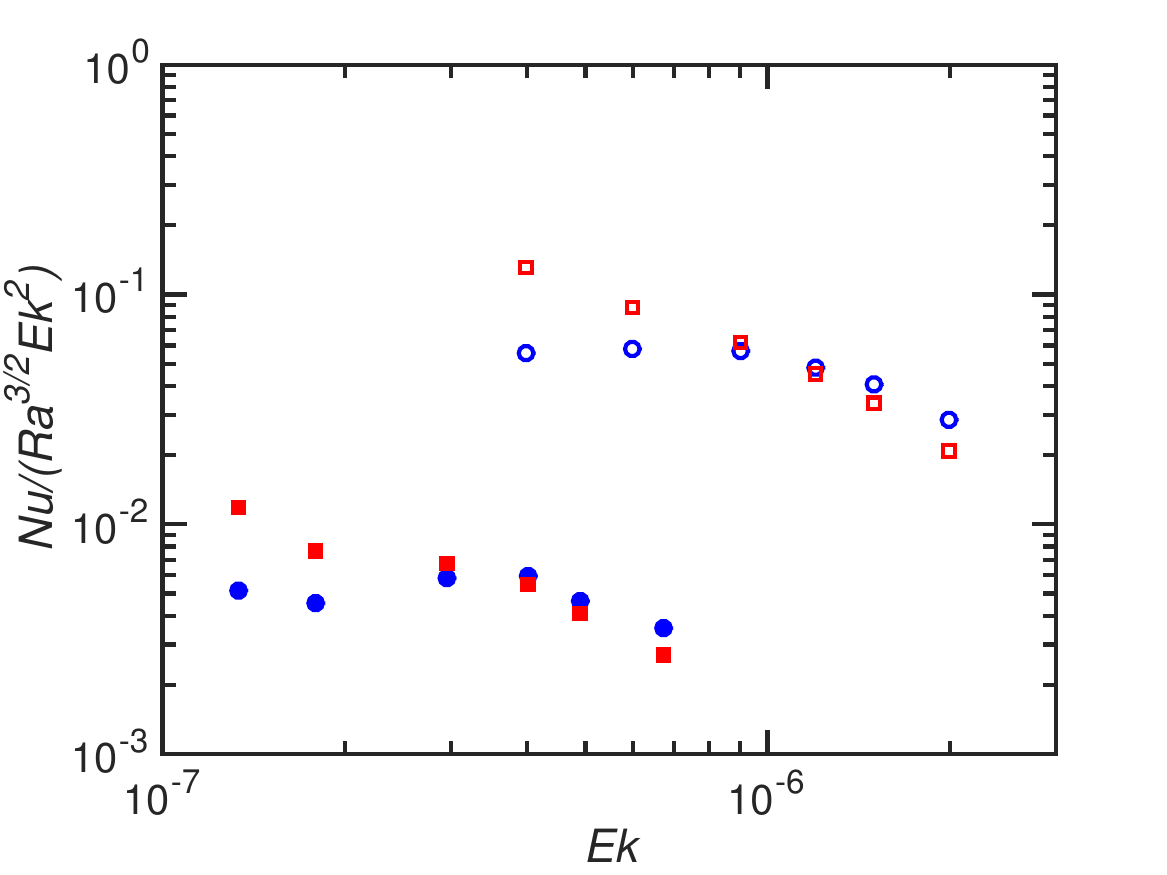}}
\caption{\label{fi:nusseltcomp} Compensated heat transfer (Nusselt number~$Nu$) as a function of the Ekman number~$Ek$ using both the  $Nu\sim Ra^3Ek^4$ (left panel) and the $Nu\sim Ra^{3/2}Ek^2$ (right panel) scaling laws proposed by \cite{ksa12} and \cite{jkrv12}.  Open symbols are for~$Ra=1\times 10^{10}$; filled symbols for~$Ra=5\times 10^{10}$, while red squares represent no-slip boundary conditions and blue circles free-slip boundary conditions. }
\end{figure}

To further quantify the scaling laws, we show in figure~\ref{fi:nusseltcomp} the compensated Nusselt number with the two scaling laws proposed by both  \cite{ksa12} and \cite{jkrv12}. Again, we can see that the $Nu\sim Ek^2$ captures well the Ekman number dependence of the free-slip simulations, but the no-slip simulations present a very different dependence. The $Nu\sim Ek^4$ scaling law can be seen to be a clear overestimate of the scaling exponent relating $Nu$ and $Ek$.

Another striking feature of this graph is that, at the same~$Ra$, there is a range for which~$Nu$ is lower for SF than for NS boundaries~\citep{sljvcrka14}. Generally, NS boundaries are expected to reduce the turbulence intensity of the flow by introducing more friction than SF plates. However, the active nature of the Ekman boundaries, present for NS but absent for SF, can affect the dynamics of the entire fluid layer, enhancing the heat transfer instead of reducing it. We will revisit these results in later sections, where further differences between NS and SF runs are revealed and interpreted.

\section{Boundary-layer and bulk dissipation\label{ch:dissipation}}
For RB convection one can derive from the Navier-Stokes equations with the Boussinesq approximation exact relations for the total dissipation of turbulent kinetic energy and thermal variance within the domain \citep{ss90}. The energy equations are obtained by taking the inner product of~$\boldsymbol{u}$ with eq.~(\ref{eq:ns}) and multiplying eq.~(\ref{eq:heat}) with~$\theta$, respectively. After applying the boundary conditions, the dissipation relations in dimensional form read:

\begin{equation}
\epsilon_u=\frac{\nu^3}{L^4}(Nu-1) RaPr^{-2}, \quad \epsilon_\theta=\kappa \frac{\Delta^2}{L^2} Nu \, ,
\label{eq:diss}
\end{equation}

\noindent where~$\epsilon_u$ is the (time- and volume-averaged) total dissipation of turbulent kinetic energy in the fluid layer and~$\epsilon_\theta$ is the total dissipation of thermal variance in the layer. These relations do not change when rotation is added: rotation only enters in the momentum equation~(\ref{eq:ns}), where we find for the Coriolis term in the energy equation that~$\boldsymbol{u \cdot}(\boldsymbol{e}_z\boldsymbol{\times u})=0$. 

The Grossmann--Lohse heat-transfer theory for non-rotating convection (see \citet{agl09} for an overview) is based on a division of the total dissipations into bulk and boundary-layer (BL) contributions. Several scaling regimes can be found depending on the dominance of dissipation in either bulk or BL regions, for both~$\epsilon_u$ and~$\epsilon_\theta$. The theoretical arguments by \citet{jkrv12} employ such a division for~$\epsilon_\theta$ (no division of~$\epsilon_u$ given that in their SF case no kinetic BLs are formed) and show that the bulk limits the overall heat transfer in geostrophic convection.

\subsection{Boundary-layer scales\label{ch:blscales}}

In this section we want to compare the distribution of dissipation for both NS and SF plates. To this end, we first need to discern between BL and bulk. Several BL scales have been already introduced in the RB literature. The thickness of the thermal BLs in the non-rotating case is well-described by assuming that the bulk is isothermal, and that the temperature drop is fully accomodated by the BLs. This leads to the definition~$\delta_{\theta,Nu}/L=1/(2Nu)$. This relation is not appropriate for rotating RB convection, given that a mean temperature drop across the bulk is sustained \citep{jlmw96}. We therefore rely on the common definition of BL thicknesses in turbulence that uses the position of the peak value of the root-mean-square of temperature fluctuations, denoted by~$\delta_\theta$.  \citet{jrgk12} found this definition to be the most appropriate one. For the kinetic (velocity) BLs we use the positions of the peak root-mean-square of horizontal velocities, marked~$\delta_{\nu}$.

\begin{figure}
\centerline{\includegraphics[width=0.49\textwidth,trim={0 0 3cm 0}, clip]{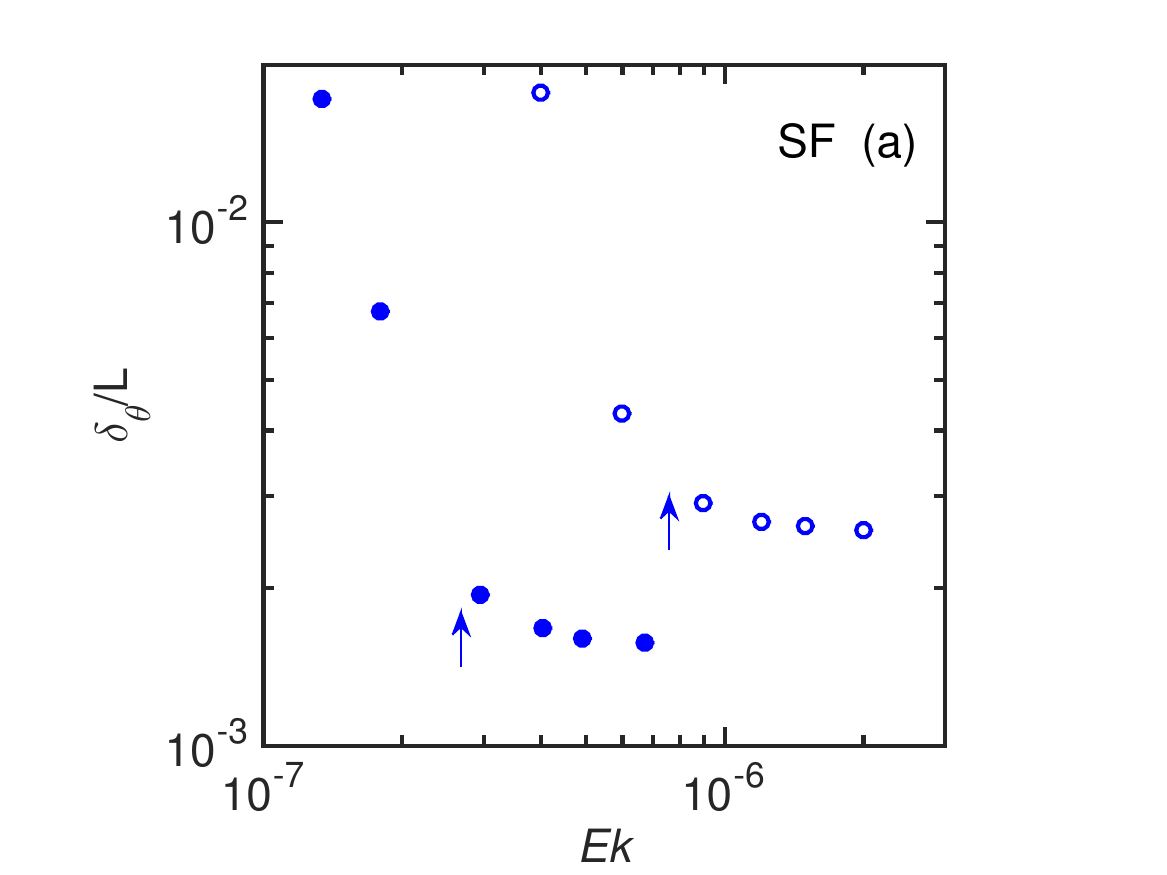}
\includegraphics[width=0.49\textwidth,,trim={0 0 3cm 0}, clip]{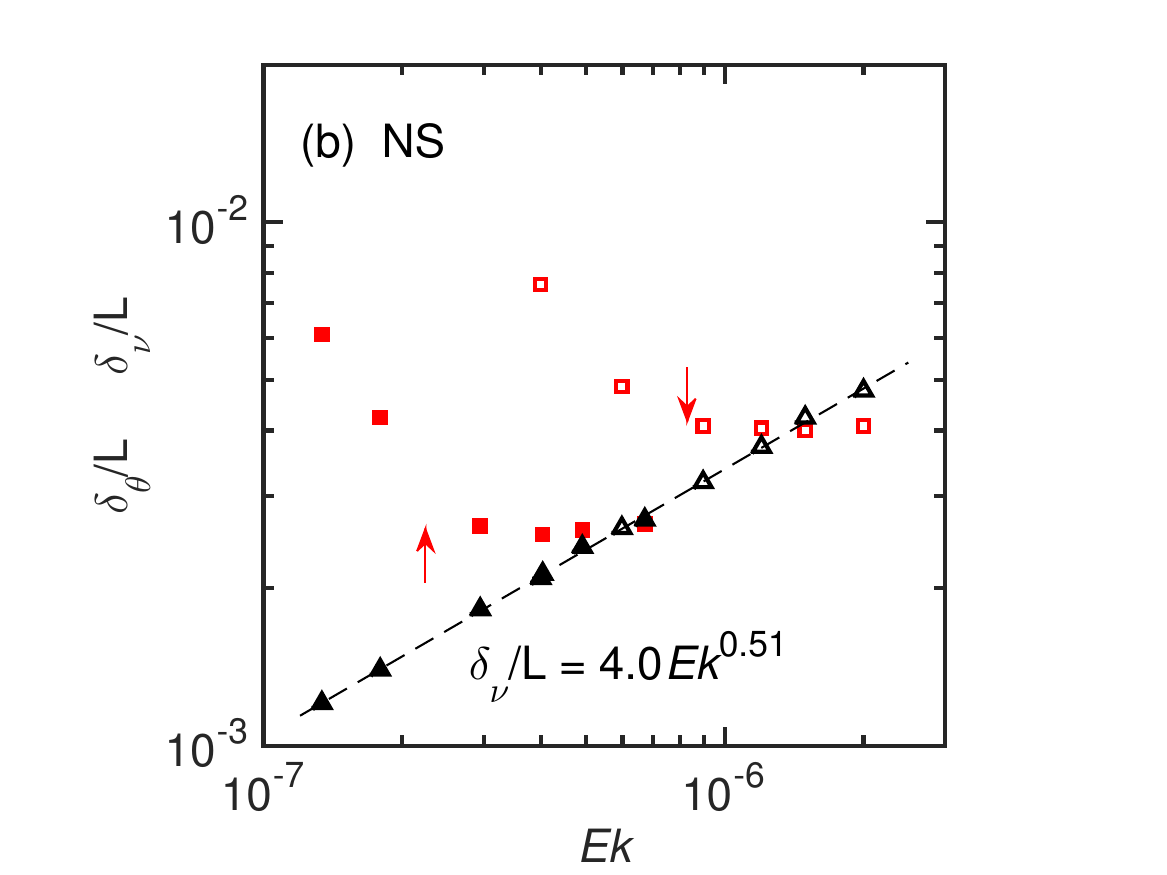}}
\caption{\label{fi:bl_scales}Boundary-layer thicknesses and their dependence on~$Ek$. The arrows indicate the transition points as inferred from these graphs, when the steeply decreasing $\delta_\theta$ begins to flatten out. (a) SF plates, thermal BL thickness ($\delta_\theta$). (b) NS plates, thermal BL thickness ($\delta_\theta$, red squares) and kinetic BL thickness ($\delta_\nu$, black triangles). A power-law fit~$\delta_\nu/L=4.0 Ek^{0.51}$ is also included (black dashed line). In both panels open symbols are for~$Ra=1\times 10^{10}$; filled symbols for~$Ra=5\times 10^{10}$.}
\end{figure}

A comparison of these BL scales is presented in figure~\ref{fi:bl_scales}. Starting from the kinetic BLs (black symbols), it is clear that they follow a single scaling, independent of~$Ra$, i.e. their thickness is \emph{exclusively} determined by~$Ek$. A power-law yields the relation~$\delta_\nu/L=4.0 Ek^{0.51}$. Within error, the slope is consistent with the prediction~$\delta_\nu\sim Ek^{1/2}$ for linear Ekman BLs \citep{g68}. It is worth noting that this scaling also matches nicely with the BL scaling laws reported by \citet{kgc10jfm}, even though the geometry (cylinder instead of periodic cube) and the Prandtl number ($Pr=6.4$ instead of $Pr=1$) are completely different. 

On the other hand, the thermal BL thicknesses do reveal some variation with~$Ek$. Before the transition the thermal BL thickness steeply decreases when~$Ek$ is increased, in contrast with to the kinetic BLs. In that $Ek$ range, the local scaling laws relating $Nu$ and $Ek$ are steeper for SF than for NS. \citet{ksnha09} and \citet{ksa12} proposed that the transition to the rotation-dominated heat-flux scaling is described by the crossing of the kinetic and thermal BL thicknesses, which happens around~$Ek=7\times 10^{-7}$ for NS and~$Ra=5\times 10^{10}$ in this case. However, the slope change in the scaling laws (figure~\ref{fi:nusselt}) is found at lower values of~$Ek$. For~$Ra=1\times 10^{10}$ a similar mismatch is observed. 

A more natural definition of the transition, based on the BL scales as plotted in figure~\ref{fi:bl_scales}, would be the evident slope change of~$\delta_\theta$ at~$Ek=8\times 10^{-7}$ for~$Ra=1\times 10^{10}$ and at~$Ek=3\times 10^{-7}$ for~$Ra=5\times 10^{10}$, which remarkably occurs around the same~$Ek$ value for both NS and SF. This transition value matches better with the slope change in the heat transfer statistics, even if they are not exactly coinciding. The transition to the geostrophic regime thus appears to be gradual; different statistics display a change in behaviour at different values of~$Ek$.

\subsection{Distribution of dissipation}

\begin{figure}
\centerline{\includegraphics[scale=0.5]{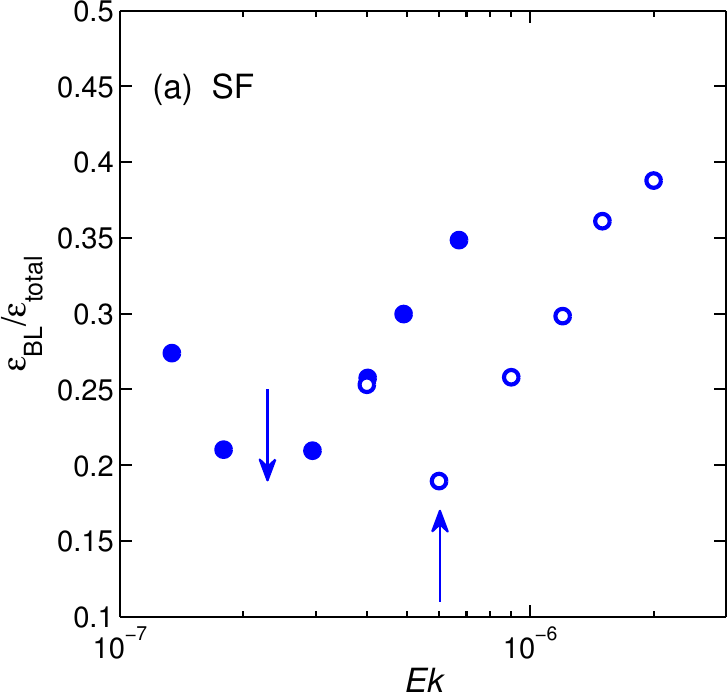}\ \ \includegraphics[scale=0.5]{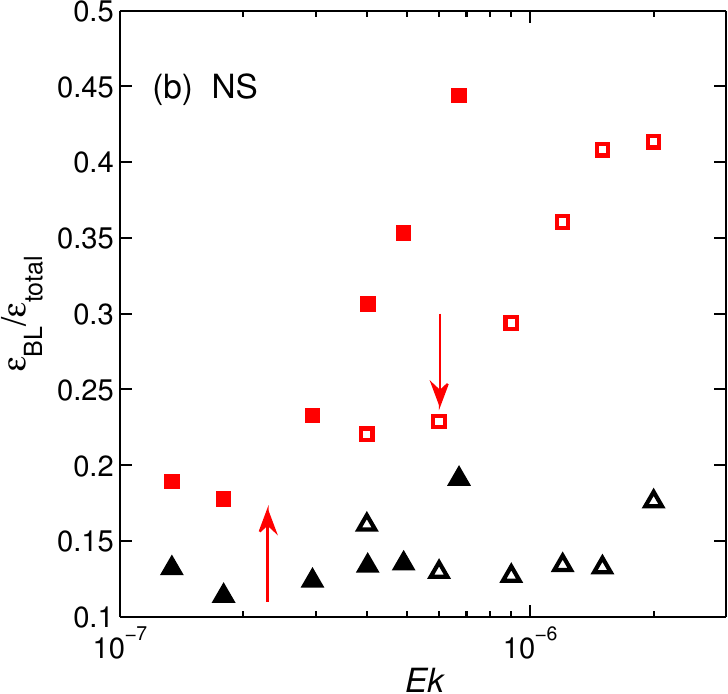}}
\caption{\label{fi:diss}Distribution of dissipations~$\epsilon_\theta$ and~$\epsilon_u$ between bulk and BL. The fraction of total dissipation located in the BL region is displayed. Open symbols are for~$Ra=1\times 10^{10}$; filled symbols for~$Ra=5\times 10^{10}$. The arrows indicate the transition points as we inferred it from these graphs. (a) SF plates, $\epsilon_\theta$. (b) NS plates, $\epsilon_\theta$ (red squares) and $\epsilon_u$ (black triangles).}
\end{figure}

Using the BL scales of~\S\ref{ch:blscales} we can now assess how the total dissipation is distributed between BL and bulk regions. This is shown in figure~\ref{fi:diss}, which confirms the picture that under rapid rotation the dissipation is mostly concentrated in the bulk. This is the case even more for~$\epsilon_u$ than for~$\epsilon_\theta$. However, we also note that the fraction of~$\epsilon_\theta$ in the BLs appears to start growing when~$Ek$ is reduced below~$\sim 5\times 10^{-7}$ ($\sim 2\times 10^{-7}$) for~$Ra=1\times 10^{10}$ ($5\times 10^{10}$), with an earlier growth appearing for SF than for NS. Looking back to figure~\ref{fi:bl_scales}, it is obvious that the thermal BLs are expanding when~$Ek$ is reduced. The larger part of the volume inside the thermal BLs along with a persistent input of thermal fluctuations from the Ekman BLs enhances the BL fraction of~$\epsilon_\theta$ at the lowest considered~$Ek$ under NS conditions. For SF plates the thermal BLs are growing even more as~$Ek$ is reduced; the increased volume of the BLs appears to be enough for a higher fraction of~$\epsilon_\theta$ there. 

Furthermore, the contribution of the kinetic BLs to the total~$\epsilon_u$ is remarkable: for~$Ra=5\times 10^{10}$, between~$Ek=1.3\times 10^{-7}$ and~$Ek=5\times 10^{-7}$ the BL thickness changes by a factor~$2$, yet the fraction of~$\epsilon_u$ in the BL remains roughly constant. This confirms that the Ekman BLs, first thought to become passive at low enough~$Ek$ \citep{nb65,jk98}, are still significantly affecting the flow dynamics \citep{sljvcrka14}.

\section{Flow phenomenology\label{ch:flow_phenom}}

\begin{figure}
\centerline{\includegraphics[scale=0.5]{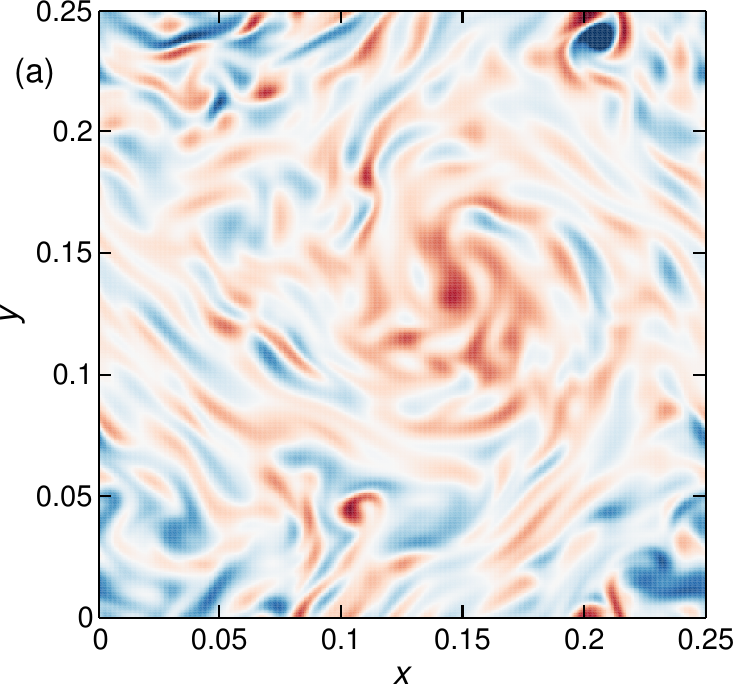}\includegraphics[scale=0.5]{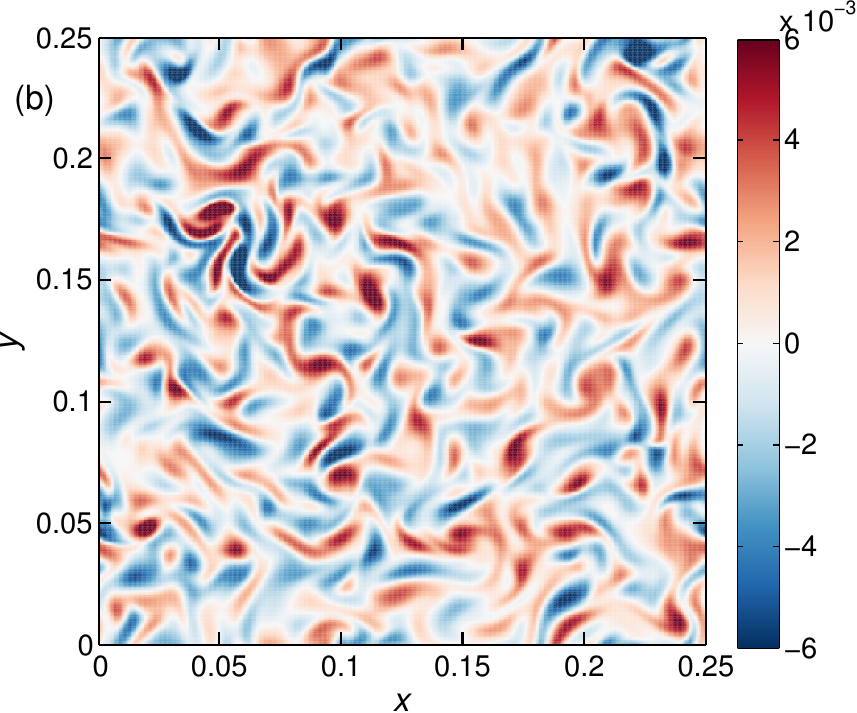}}
\caption{\label{fi:snaps}Snapshot at midheight ($z=0.5$) of the vertical vorticity from runs at~$Ra=5\times 10^{10}$ and~$Ek=1.34\times 10^{-7}$. (a) SF plates. (b) NS plates. Red is positive (cyclonic) vorticity, while blue is negative (anticyclonic) vorticity. Both plots have the same colour scale.}
\end{figure}

A remarkable phenomenological change upon entering the geostrophic regime is the disappearance of convective Taylor columns and plumes \citep{sjkw06,jrgk12,sljvcrka14}. These vortical structures have been frequently reported ever since the first observation in turbulent rotating RB flow by \citet{r69}. In the geostrophic regime, however, such coherent structures seem to be absent. The boundary conditions largely determine the flow phenomenology. For SF plates large barotropic vortices can be formed under the influence of an inverse energy cascade \citep{rjkw14,fsp14,ghj14}, eventually growing to the scale of the domain. For NS plates such a condensate is not formed; we compare the phenomenology in two snapshots shown in figure~\ref{fi:snaps}, which depict the spatial distribution of vertical vorticity~$\omega_z=\partial_x u_y-\partial_y u_x$ in a horizontal cross-section at midheight. Panel (a) clearly reveals the formation of a large cyclonic vortex in the top right, while the bottom-right part and its periodic continuation on opposite sides hint at the formation of an anticyclonic vortex. Note that this flow is still slowly evolving over time \citep{rjkw14,fsp14,ghj14}. Panel (b) shows no condensate vortices. Instead, a fluctuating state is found without large-scale long-lived coherent structure. Ekman pumping, present only in the case of NS boundary conditions, can thus be a source of small-scale fluctuations that prevent condensation into large-scale vortices.

\subsection{Relation with spectral energy transfer}

In previous works \citep{rjkw14,fsp14} the energy transfer as a function of wavenumber has been considered to indicate the presence of an inverse energy cascade. Given that figure~\ref{fi:snaps} reveals such a significant difference in flow phenomenology between SF and NS, we anticipate that the spectral energy transfer must also be considerably different. Following \citet{fsp14}, we define the spectral energy equation as
\begin{equation}
\frac{d E(K)}{dt}=\sum_Q T(Q,K)-D(K)+F(K) \, ,
\end{equation}

\noindent which gives the temporal evolution of the energy~$E(K)$ as a function of horizontal wavenumber~$K$ in terms of the energy transfer~$T(Q,K)$ from wavenumber shell~$Q$ to shell~$K$, the dissipation~$D(K)$ and the buoyant forcing~$F(K)$. We are particularly interested in the transfer term~$T(Q,K)$; which we evaluate and average vertically over the entire computational domain minus the BLs. We note that this quantity is antisymmetric by definition, i.e.~$T(K_1,K_2)=-T(K_2,K_1)$ and~$T(K_1,K_1)=0$ for any two wavenumbers~$K_1$ and~$K_2$.

\begin{figure}
\centerline{\includegraphics[scale=0.5]{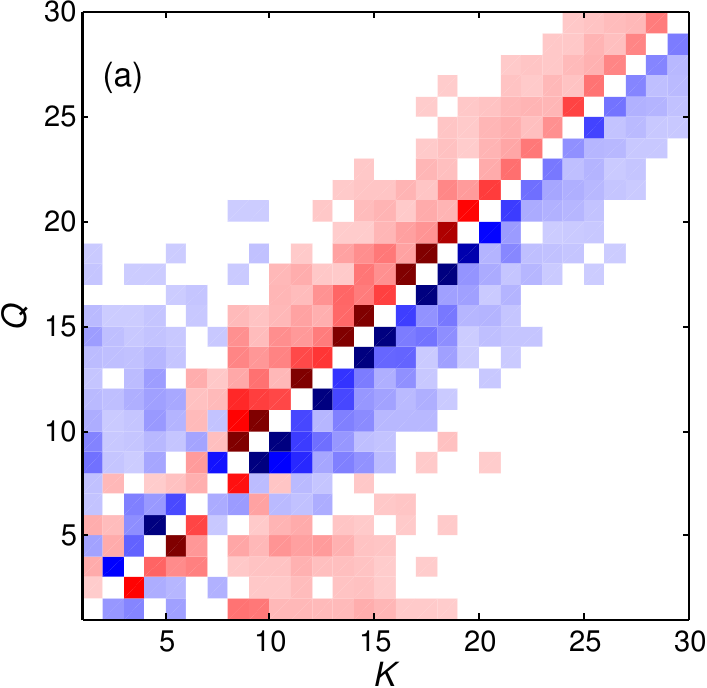}\includegraphics[scale=0.5]{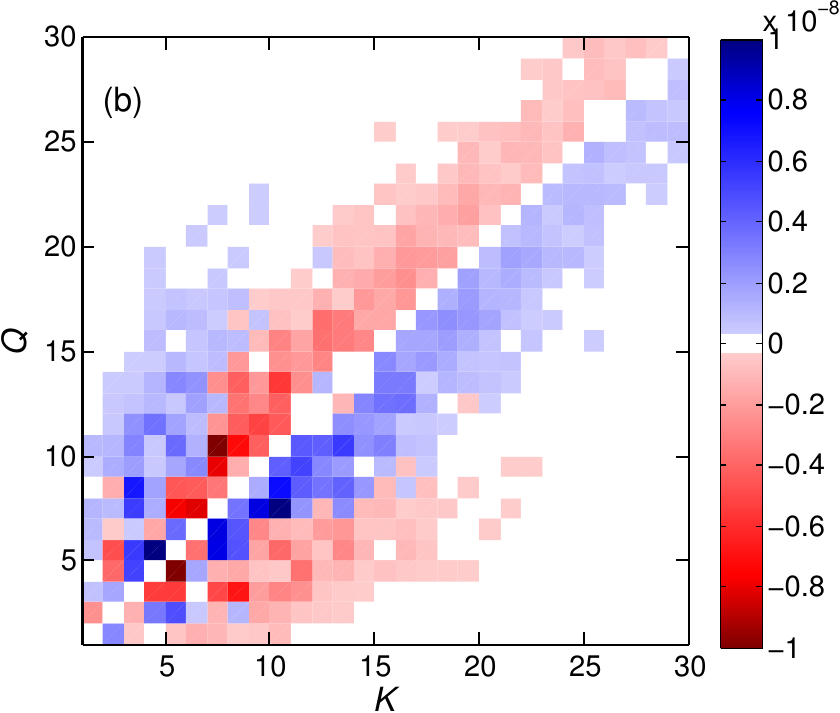}}
\caption{\label{fi:spectral_energy}Spectral energy transfer~$T(Q,K)$ between wavenumber shells~$Q$ and~$K$. Positive values indicate that energy is taken from shell~$Q$ and transferred to shell~$K$, negative values imply that~$Q$ receives energy from~$K$. Panel (a) is for SF and panel (b) for NS. Both plots have the same colour scale.}
\end{figure}

The spectral transfer at~$Ra=5\times 10^{10}$ and~$Ek=1.34\times 10^{-7}$ is depicted in figure~\ref{fi:spectral_energy}, for both SF and NS conditions. The SF picture compares favourably to the earlier studies by \citet{fsp14} and \citet{rjkw14}: a strong localised exchange of energy between neighbouring wavenumbers across the diagonal, i.e., a certain mode~$P$ interacts predominantly with its neighbouring modes~$P-1$ and~$P+1$. The sign of the transfer reveals that this part of the cascade is direct: Energy is transferred to larger wavenumbers. At low~$K\lesssim 7$ and higher~$Q\gtrsim 7$, a range of inverse transfer is found where small-$K$ modes receive energy from modes~$Q>K$. The input of energy into the large-scale vortex of figure~\ref{fi:snaps}(a) can be recognised as an interaction of the smallest wavenumbers receiving energy from a broad range of higher-order wavenumbers. 

For the NS case (figure~\ref{fi:spectral_energy}(b)), the picture changes. In particular, the `staircase' of energy transfer along the diagonal, which is prominent for SF, is not as strongly present for the larger wavenumbers. Instead, the interactions between modes are less localised, meaning that the interactions are spread more and mode combinations farther from the diagonal are transferring energy. A direct cascade is formed along the diagonal for~$K,Q\gtrsim 5$. Curiously, some signs of an inverse cascade remain: (i) for~$K\le 5$, and (ii) transfers from modes around~$Q\approx 5$ to~$K\approx 10$, which are non-localised. The absence of a large-scale structure can be explaned by the fact that the lowest wavenumber is not actively receiving energy.

\subsection{Relation with mean temperature gradient}
\begin{figure}
\centerline{\includegraphics[scale=0.5]{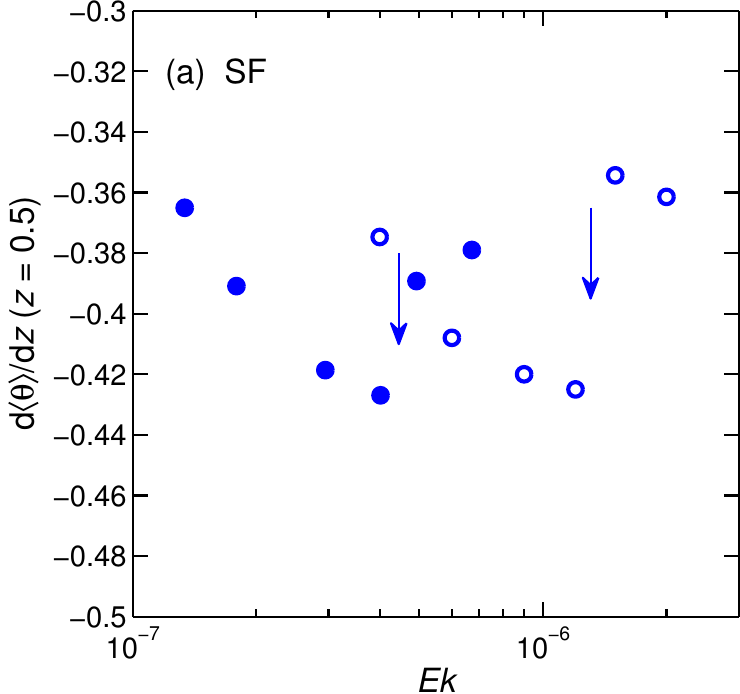}\ \ \includegraphics[scale=0.5]{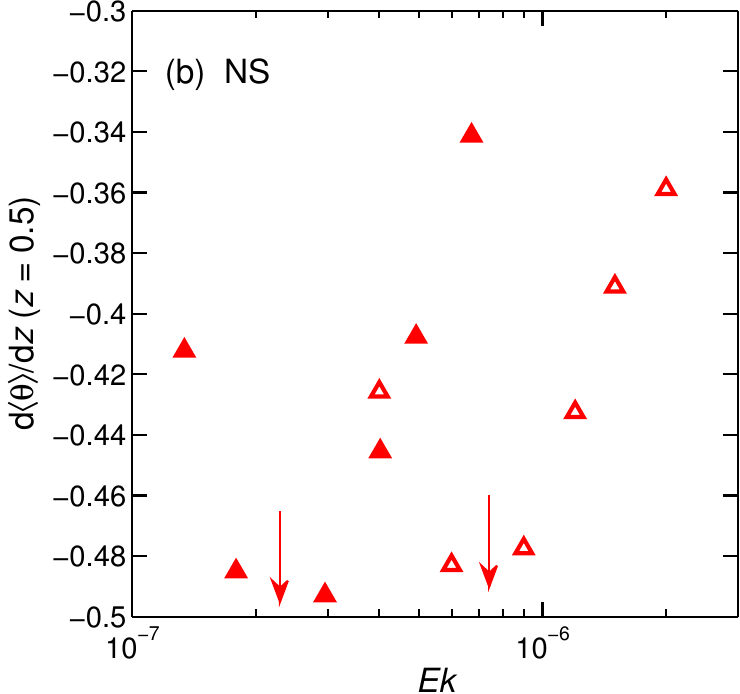}}
\caption{\label{fi:tgrad}Mean temperature gradient at midplane. Open symbols are for~$Ra=1\times 10^{10}$; filled symbols for~$Ra=5\times 10^{10}$. The arrows indicate the transition points as we inferred it from these graphs. Panel (a) is for SF and panel (b) for NS.}
\end{figure}

The characteristic flow phenomenology of rotating RB convection has been related to the occurrence (and strength) of the persistent mean temperature gradient across the bulk \citep{jlmw96}, unlike the statistically isothermal bulk of non-rotating RB. The origin of this temperature gradient has been proposed to be increased by lateral mixing, induced by interactions of like-signed vortical plumes. As in the geostrophic regime the flow phenomenology is altered (c.f. figure \ref{fi:snaps}), we can expect that this also affects the strength of the mean temperature gradient. Figure~\ref{fi:tgrad} shows the mean temperature gradient as a function of $Ek$. In the rotation-affected RB, lateral mixing being stronger that vertical mixing leads to mean temperature gradients as large as~$-0.5$ for NS and $-0.4$ for SF. However, upon entering the geostrophic regime by further reducing~$Ek$, the magnitude of the gradient is gradually diminished. We thus expect that in the geostrophic regime the mixing can become slightly more three-dimensional again.

This behaviour is consistent with previous simulations of the asymptotic equations \citep{jrgk12}, which predict that the geo\-stroph\-ic regime indeed still has a mean temperature gradient, but less steep than when coherent vortical plumes are present. Finally, even though the behaviour of the mean temperature gradient is qualitatively similar for SF and NS plates, the location of the minimum temperature gradient, which could be taken as an additional indicator for the transition, is certainly not coinciding between the two cases.

\section{Discussion\label{ch:discussion}}

In the previous sections we have compared the transition to the geostrophic regime of turbulent rotating Rayleigh--B\'enard convection between stress-free (SF) and no-slip (NS) boundary conditions on the horizontal plates. From the current results, it seems clear that the nature of the bulk turbulence is extremely dependent on the boundary conditions. Nevertheless, both types of boundary conditions display a transition in a similar range of Ekman numbers around~$Ek=9\times 10^{-7}$ ($3\times 10^{-7}$) for~$Ra=1\times 10^{10}$ ($5\times 10^{10}$). This transition is found to be gradual, unlike other transitions in rotating Rayleigh--B\'enard, such as those reported in confined geometries at higher~$Ek$ \citep{szcal09}. Many diagnostic signs of flow transition can be found near the onset of the geostrophic regime: The scaling with~$Ek$ of many quantities including Nusselt number, thermal BL thickness, bulk--BL distribution of dissipation rates as well as the mid-plane mean temperature gradient show a changing behaviour. We do not expect this list to be exhaustive. All quantities show a transition centered at a specific~$Ek$, so that the full range of changes covers at least half a decade in~$Ek$. In particular, at~$Ra=1\times 10^{10}$ we inferred transitions in various statistics in the range~$6\times 10^{-7}\lesssim Ek\lesssim 1.3\times 10^{-6}$; at~$Ra=5\times 10^{10}$ in the range~$2.4\times 10^{-7}\lesssim Ek\lesssim 4.5\times 10^{10}$. These ranges are the same for SF and NS plates, however, individual statistics display transitions at different~$Ek$ when comparing the two boundary conditions. So it appears to be all but impossible to define a single criterion to distinguish the rotation-dominated geostrophic regime from the rotation-affected regime characterised by vortical plumes or columns, as indeed the flow may be transitioning, but the different diagnostic quantities may be sensitive slightly before or slightly after the transition.

Regarding the actual nature of the transition, it is quite remarkable that the SF and NS transition ranges are coinciding in $Ek$ for similar $Ra$. This would suggest a common origin. One of two candidates suggested in the literature may be the cause (or a combination of both): either marginal (in)stability of the thermal BL, as suggested by \citet{ksa12}, which leads to a theoretical scaling~$Nu\sim Ra^3Ek^4$; or a change in the bulk dynamics where plumes cannot enter the stiff geostrophic bulk that throttles the heat transport, suggested by \citet{jkrv12}, which gives a scaling~$Nu\sim Ra^{3/2}Ek^2$. Both arguments could in principle be independent of the velocity boundary conditions, given that either the thermal BLs or the bulk flow away from the BLs is involved. We find mostly evidence supporting the \citet{jkrv12} mechanism, but it is certainly not conclusive:

\begin{itemize}
\item The geostrophic Nusselt number scaling of figure~\ref{fi:nusselt} matches fairly with the \citet{jkrv12} scaling for SF plates, but not for NS. The scaling proposed by~\citet{ksa12} does not match with the current results for either boundary condition. In line with the recent experimental results of \citet{csrgka15}, it is becoming clear that the heat-transfer scaling exponent~$\beta$ for $Nu\sim Ra^\beta$ measured at constant~$Ek$ is not the same for all~$Ek$; equivalently, the exponent~$\gamma$ for~$Nu\sim Ek^\gamma$ at constant~$Ra$ will take different values for different~$Ra$.
\item For the BL thickness (figure~\ref{fi:bl_scales}) we find a change in scaling at~$Ek=8\times 10^{-7}$ ($3\times 10^{-7}$) at~$Ra=1\times 10^{10}$ ($5\times 10^{10}$) for both SF and NS, with steeper scaling with~$Ek$ below the transition. The transition does not coincide with the crossing of the kinetic and thermal BL thicknesses, which is the criterion proposed by \citet{ksnha09} and \citet{ksa12} to describe its origin. The sharp change of scaling indicates changes in the structure of the thermal BL, which could be due to the crossing of the marginal stability criterion for the BL. However, the corresponding limit of validity~$Ra\lesssim Ek^{-3/2}$ for the argument \citep{ksa12} would at the current~$Ra=5\times 10^{10}$ predict a transition at~$Ek\approx 7\times 10^{-8}$, at significantly smaller~$Ek$ than we observe.
\item Finally, the spatial distribution of dissipation of both turbulent kinetic energy ($\epsilon_\theta$) and thermal variance ($\epsilon_u$) between bulk and BL (figure~\ref{fi:diss}) reveals that most of the dissipation is found in the bulk, even more so for~$\epsilon_u$ than for~$\epsilon_\theta$. The fractional distribution between bulk and BL reveals a slope change at~$Ek\approx 6\times 10^{-7}$ ($2\times 10^{-7}$) at~$Ra=1\times 10^{10}$ ($5\times 10^{10}$) for both SF and NS. 
\end{itemize}

For SF plates the \citet{jkrv12} arguments fit best with our findings. However, the case of NS plates requires a different description given the presence of Ekman layers that are significantly affecting the flow dynamics in the entire fluid layer. We presently cannot give a theoretical description now, but we expect that these results, together with the recent findings by \citet{sljvcrka14} and \citet{csrgka15}, can form a starting point for theories of no-slip geophysical convection.

In conclusion, it has become apparent in the last few years that the Ekman layers remain a decisive and active part of geostrophic convection with no-slip plates, in spite of their diminishing thickness. We have compared the transition to the geostrophic regime between no-slip and stress-free boundaries. Both undergo a transition, at roughly the same Ekman number, but the scaling laws for heat-transfer on both sides of the transition are strongly dependent on the boundary conditions.  The physical picture of geostrophic convection is not fully complete, especially for no-slip plates.

\acknowledgments

We would like to thank S. Grossmann, C. Sun, and Y. Yang for various stimulating discussions. We acknowledge FOM, an ERC Advanced Grant, the PRACE resource Hermit based in Stuttgart at HLRS, and NWO for the use of Cartesius under Grant No. SH-202.

\bibliographystyle{jfm}
\bibliography{kunnen}
\end{document}